\documentclass[submission,copyright]{eptcs}
\providecommand{\event}{
ETAPS Workshop on Verification and Program Transformation (VPT 2021)} 

\usepackage{breakurl}             
\usepackage{underscore}           
\usepackage{amsmath}
\usepackage{mathtools}
\usepackage{array}
\usepackage{multirow}
\usepackage{booktabs}

\usepackage{paralist}

\usepackage{tikz}
\usepackage{verbatimbox}
\usetikzlibrary{matrix}

\usepackage{todonotes}
\usepackage{comment}
\usepackage{hyperref}

\usepackage{caption}
\usepackage{subcaption}

\usepackage{ascmac}	

\title{An Inversion Tool for Conditional Term Rewriting Systems\\ -- A Case Study of Ackermann Inversion}

\author{Maria Bendix Mikkelsen\thanks{Thanks for support to Zonta International, Club of Copenhagen.} 
\qquad Robert Gl\"uck 
\institute{DIKU, University of Copenhagen, Denmark 
\email{mbm@di.ku.dk, glueck@acm.org}}
\and
Maja H.\ Kirkeby
\institute{Roskilde University, Denmark
\email{kirkebym@acm.org}}}

\def\titlerunning{An Inversion Tool for Conditional Term Rewriting Systems -- A Case Study of Ackermann Inversion}
\def\authorrunning{M.B.\ Mikkelsen, R.\ Gl\"uck, M.H.\ Kirkeby}

\begin{document}
\maketitle

\begin{abstract}
We report on an inversion tool for a class of oriented conditional constructor term rewriting systems. Four well-behaved rule inverters ranging from trivial to full, partial and semi-inverters are included.
Conditional term rewriting systems are theoretically well founded and can model functional and non-functional rewrite relations.
We illustrate the inversion by
experiments with full and partial inversions of the Ackermann function. The case study demonstrates, among others, that polyvariant inversion and input-output set propagation can
reduce the search space of the generated inverse systems.

\end{abstract}

\section{Introduction}
\label{sec:introduction}

Program inversion is one of the fundamental transformations that can be performed on programs~\cite{GlueckKlimov:94:LMMC}. Although function inversion is an important concept in mathematics, program inversion has received little attention in computer science.
In this paper,
we report on a tool implementation
of an
inversion framework~\cite{KirGlu:20}
and on some computer experiments within the framework. 
The implementation includes four
well-behaved rule inverters
ranging from trivial to full, partial and semi-inverters, several of which have been studied in the literature~\cite{KirGlu:19,Nishida2004,Nishida2005}.
The generic inversion algorithm
used by the tool
was proven to produce the correct result for all well-behaved rule inverters~\cite{KirGlu:20}.
The tool reads the standard notation of the
established
confluence competition  (\href{http://project-coco.uibk.ac.at/}{COCO}), making it compatible with
other
term rewriting tools.
The Haskell implementation
is designed as an open system for experimental and educational purposes
that can be extended
with further well-behaved
rule inverters.

In particular, we illustrate the use of the tool by repeating A.Y.\ Romanenko's three experiments with full and partial inversions of the Ackermann function~\cite{Romanenko1988,Romanenko1991}.
His inversion algorithm, inspired by Turchin~\cite{Turchin:86},
inverts programs written in a Refal extension, Refal-R~\cite{Romanenko1991}, which is a functional-logic language, whereas our tool uses a subclass of oriented
 conditional constructor term rewriting systems\footnote{CCSs are also referred to as pure constructor CTRS~\cite{Nagashima2012}.} (CCSs)~\cite{Bezem2003,Ohlebusch2002}. 
 Conditional term rewriting systems 
 are theoretically well founded and can model a wide range of language paradigms, e.g., reversible, functional, and declarative  languages.
 
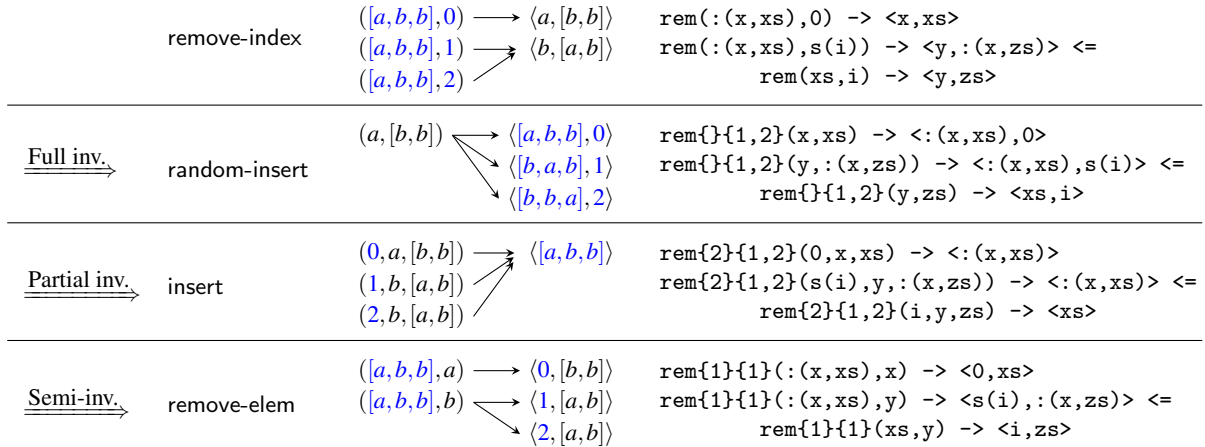
\begin{figure}[ht]
{\footnotesize
\centering
\begin{tabular}{m{1.5cm} m{1.9cm} m{3.8cm} m{7cm} }
&\textsf{remove-index} \vspace{-0.1cm}

& 
\begin{tikzpicture}
  \matrix (m) [matrix of math nodes,row sep=-3pt,column sep=2em,minimum width=3em,nodes in empty cells]
  {
   	{( {\color{blue}[a,b,b]},{\color{blue}0})} & {\langle a,[b,b]\rangle} \\
	{({\color{blue}[a,b,b]},{\color{blue} 1})} & {\langle b,[a,b]\rangle} \\
	{( {\color{blue}[a,b,b]},{\color{blue}2})} & \\
     };
     
  \path[-stealth]
  
(m-1-1.east) edge  
(m-1-2.west)
(m-2-1.east) edge[shorten >=0.5mm]   (m-2-2.west) 
(m-3-1.east) edge[shorten >=0.75mm]   (m-2-2.west) 
    ;
\end{tikzpicture}
&
{
\begin{verbatim}
rem(:(x,xs),0) -> <x,xs>                   
rem(:(x,xs),s(i)) -> <y,:(x,zs)> <=        
        rem(xs,i) -> <y,zs> \end{verbatim}
}
\\[-6pt] \cline{1-4}
\vspace{-0.55cm}
{$\xRightarrow{\text{\footnotesize{Full inv.}}}$} &
\textsf{random-insert} 
 & 
\begin{tikzpicture}

  \matrix (m) [matrix of math nodes,row sep=-3pt,column sep=2em,minimum width=3em,nodes in empty cells]
  { 	{( a,[b,b])} & {\langle {\color{blue}[a,b,b]},{\color{blue}0}\rangle} \\
			& {\langle {\color{blue}[b,a,b]},{\color{blue}1}\rangle} \\
			& {\langle {\color{blue}[b,b,a]},{\color{blue}2}\rangle} \\
     };
     
  {  \path[-stealth]

(m-1-1.east) edge 
 (m-1-2.west)
(m-1-1.east) edge   (m-2-2.west)
(m-1-1.east) edge   (m-3-2.west)
  ;}
\end{tikzpicture}
& 
{\footnotesize
\begin{verbatim}
rem{}{1,2}(x,xs) -> <:(x,xs),0>
rem{}{1,2}(y,:(x,zs)) -> <:(x,xs),s(i)> <=
        rem{}{1,2}(y,zs) -> <xs,i>
\end{verbatim}
}
\\[-6pt] \cline{1-4}
{$\xRightarrow{\text{\footnotesize{Partial inv.}}}$ }&
\textsf{insert} \vspace{-0.1cm}
&
\begin{tikzpicture}
  \matrix (m) [matrix of math nodes,row sep=-3pt,column sep=2em,minimum width=3em,nodes in empty cells, column 1/.style={anchor=base west}]
  {	{( {\color{blue}0},a,[b,b])} &  {\langle {\color{blue}[a,b,b]} \rangle} \\
	{( {\color{blue}1},b,[a,b])} & \\ 
	{( {\color{blue}2},b,[a,b])} & \\ 
     };
     
  { \path[-stealth, shorten >=0.75mm] 
(m-1-1.east) edge  
(m-1-2.west)
(m-2-1.east) edge[shorten >=1mm]   (m-1-2.west)
(m-3-1.east) edge[shorten >=1.5mm]   (m-1-2.west)
    ;}
\end{tikzpicture} 
&
{\footnotesize
\begin{verbatim}
rem{2}{1,2}(0,x,xs) -> <:(x,xs)>
rem{2}{1,2}(s(i),y,:(x,zs)) -> <:(x,xs)> <=
        rem{2}{1,2}(i,y,zs) -> <xs>
\end{verbatim}
}
\\[-6pt] \cline{1-4}
{ $\xRightarrow{\text{\footnotesize{Semi-inv.}}}$}&
\textsf{remove-elem} \vspace{-0.1cm}
&
\begin{tikzpicture}
  \matrix (m) [matrix of math nodes,row sep=-3pt,column sep=2em,minimum width=3em,nodes in empty cells, column 1/.style={anchor=base west}]
  {
        {( {\color{blue}[a,b,b]}, a )} & {\langle {\color{blue}0}, [b,b] \rangle} \\
	{( {\color{blue}[a,b,b]}, b)} & {\langle {\color{blue}1},  [a,b]\rangle} \\
				 & {\langle {\color{blue}2}, [a,b]\rangle} \\
     };
     
  { \path[-stealth]
  (m-1-1.east) 	edge  
  (m-1-2.west)
 (m-2-1.east) 	edge   (m-2-2.west)
			edge   (m-3-2.west);}
\end{tikzpicture} &
{\footnotesize
\begin{verbatim}
rem{1}{1}(:(x,xs),x) -> <0,xs>
rem{1}{1}(:(x,xs),y) -> <s(i),:(x,zs)> <=
        rem{1}{1}(xs,y) -> <i,zs>
\end{verbatim}
} \\[-8pt]
\end{tabular}
}
\caption{Full, partial and semi-inversion of the remove-index function \texttt{rem}.}
\label{fig:rem-inv}
\end{figure}
Let us illustrate our tool with three kinds of inversions
of a simple \textsf{remove-index} function, \texttt{rem} (Figure~\ref{fig:rem-inv}). 
Given a list and a unary number $n$, it returns the $n$th element of the list and the list without the removed element.
\texttt{rem} is defined by two rewrite rules: the first rule defines the base case where cons~({\footnotesize\verb+:+}) is written in prefix notation and the two outputs are tupled ({\footnotesize \texttt{<$\cdot$>}}). 
The second rule contains a so-called condition after the separator ({\footnotesize \texttt{<=}}) that can be read as a recursion.
The input-output relation specified by \texttt{rem} is exemplified
by the list $[a,b,b]$ and the indices 0, 1, and 2 (inputs are marked in blue).

Usually, we consider program inversion as \emph{full inversion} that swaps a program's entire input and output.
In our tool, the new directionality of the desired program is specified by input and output index sets (\emph{io-sets}). The user can select the input and output arguments that become the input arguments of the inverse program, so this technique is very general.
Full inversion always has an empty input index set~$I$ and an
output index set~$O$ containing all indices of the outputs. 

Full inversion of \texttt{rem} yields a program \verb+rem{}{1,2}+ whose non-functional
input-output
relation specifies the insertion of an element into a list at an arbitrary position. The updated lists and the corresponding positions
are the output.
The name
\verb+rem{}{1,2}+
indicates that none of \texttt{rem}'s inputs (\verb+{}+) and all of \texttt{rem}'s outputs (\verb+{1,2}+) are the new inputs.

The full inverse of a non-injective function specifies a non-functional relation. Thus, program inversion
does \emph{not} respect language paradigms, and this is one of the inherent difficulties when performing program inversion for a functional language.
The inverted rules do not always define a functional relation because they
may have
overlapping left-hand sides or extra variables.
The non-functional relation \textsf{random-insert} of \verb+rem{}{1,2}+ is induced by two overlapping rules.

\emph{Partial inversion} swaps parts of the input and the entire output.
\verb+rem{2}{1,2}+ is a partial inverse of \verb+rem+ where the original list is swapped with the entire output. It defines the insertion of an element at a position $n$ in a list, i.e., the functional relation \textsf{insert}.
\emph{Semi-inversion}, the most general form of inversion, can swap any part of the input and output.
\verb+rem{1}{1}+ is a semi-inversion of \verb+rem+ where the position and the element are swapped, i.e., the non-functional relation \textsf{remove-elem}.
While we obtain two programs for the price of one by full inversion, we can obtain several programs by partial and semi-inversion.\\[-0.5ex]

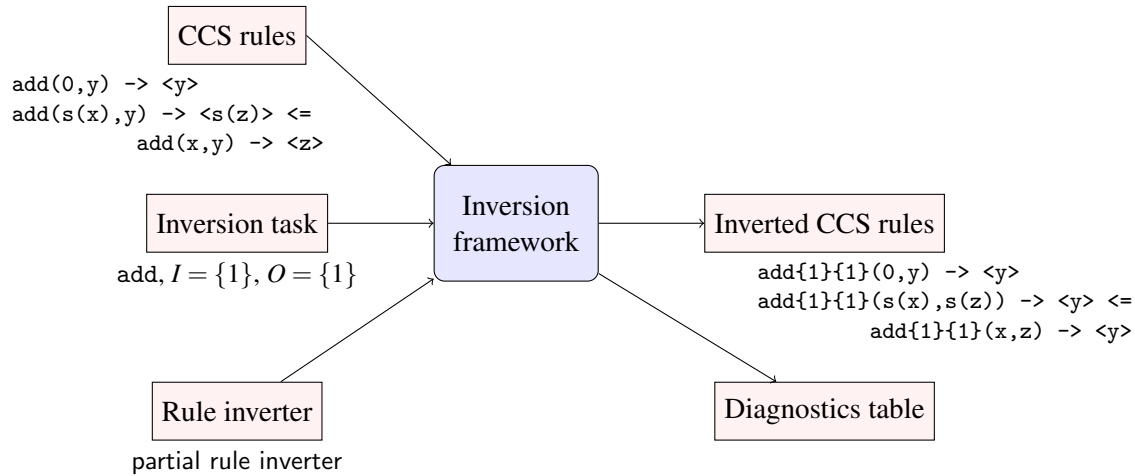
\begin{figure}[t]
\tikzstyle{block} = [rectangle, draw, fill=blue!10, 
    text width=5em, text centered, rounded corners, minimum height=4em]
\tikzstyle{line} = [draw, ->, -latex']
\tikzstyle{cloud} = [draw, fill=red!5, node distance=2.5cm, minimum height=2em]
\begin{center}

\begin{myverbbox}[\footnotesize]{\ccsadd}
add(0,y) -> <y>
add(s(x),y) -> <s(z)> <= 
          add(x,y) -> <z>
\end{myverbbox}

\begin{myverbbox}[\footnotesize]{\ccsaddinv}
add{1}{1}(0,y) -> <y>
add{1}{1}(s(x),s(z)) -> <y> <= 
         add{1}{1}(x,z) -> <y>
\end{myverbbox}  

\begin{tikzpicture}[node distance = 2cm, auto]
    \node [block] (init) {Inversion framework};

    \node [cloud, left of=init, anchor=east] (task) {\text{Inversion task}};
    \node [cloud, below of=task
    ] (ruleinv) {\text{Rule inverter}};
    \node [cloud, above of=task
    ] (ccs) {\text{CCS rules}};
    
    \node at (ccs.south west) [anchor=north
    ] (ccs') {
    {\ccsadd}
    };
    \node at (ruleinv.south) [anchor=north] (ruleinv') {\small\textsf{partial rule inverter}};
    \node [cloud, right of=init,anchor=west] (invs) {\text{Inverted CCS rules}};
    \node [cloud, below of=invs,xshift=0.0mm] (diag) {\text{Diagnostics table}};
    \node at (task.south) [anchor=north
    ] (ruleinv') {\small\texttt{add}, $I = \{1\}$, $O = \{1\}$};
    \node at (invs.south east) [anchor=north
    ]
    (invs') 
    {\ccsaddinv};
    
    \path [draw,->] (ccs.east) -- (init);
    \path [draw,->] (ruleinv) -- (init);
    \path [draw,->] (task.east) -- (init);
    \path [draw,->] (init) -- (invs);
    \path [draw,->] (init) -- (diag);
\end{tikzpicture}
\end{center}
\caption{
The tool 
illustrated with the partial inversion of  \texttt{add} defining the addition of unary numbers.}
\label{fig:tool}
\end{figure}

\noindent The contribution of the work reported here
is a complete implementation of the generic inversion algorithm, together with four well-behaved rule inverters.
The system is available for experimental (source code) and educational purposes (web system).
We then report on the case study of Ackermann inversion repeating three experiments by A.Y.\ Romanenko and compare the results and provide measures of the 
rewrite steps and function calls. We are not aware of other investigations of Romanenko's experiments.

We begin by giving an overview of the tool (Section~\ref{sec:inversiontool}) followed by the case study of Ackermann inversion (Section~\ref{sec:ackermann}).
This paper, the tool implementation itself, and the paper on the inversion framework are intended to complement each other. 
They are intended to be used together.
For more details on the generic inversion algorithm and the rule inverters, interested readers are therefore referred to~\cite{KirGlu:20}.

\section{Inversion Tool}
\label{sec:inversiontool}

The tool is implemented in Haskell\footnote{The Glorious Glasgow Haskell Compilation System, version 8.10.4}, and we provide both an online web-based version\footnote{\url{https://topps.di.ku.dk/pirc/inversion-tool}} and the source code\footnote{\url{https://github.com/pirc-src/inversion-tool}}.
We demonstrate how to
partially invert a function \texttt{add}, which defines the addition of two unary numbers,
to obtain \texttt{add\{1\}\{1\}}, which defines the subtraction of two unary numbers.
%
As illustrated in Figure~\ref{fig:tool}, the \textit{Inversion Framework} requires three inputs:
\begin{inparaenum}[(i)]
\item the original \textit{CCS rules} with the \texttt{add}-rules,
\item the \textit{Inversion task}:
\texttt{add} with io-set $I=\{1\}$ and $O=\{1\}$, and
\item an indication of which well-behaved \textit{Rule inverter} the tool should apply, here, the partial rule inverter.
\end{inparaenum}
The inversion framework provides two outputs:
\begin{inparaenum}[(i)]
    \item the \textit{Inverted CCS rules}, containing the \texttt{add\{1\}\{1\}}-rules defining subtraction, and 
    \item a \textit{Diagnostics table} with an overview of the systems' paradigm characteristics. 
\end{inparaenum}
Because the program inversion does not respect language paradigms, it is useful that the tool also provides an analysis of the programs' paradigm characteristics; see~\cite[Fig.2]{KirGlu:20} for definitions and interrelations.
\begin{figure}[t]
\begin{center}
\includegraphics[width=\textwidth]{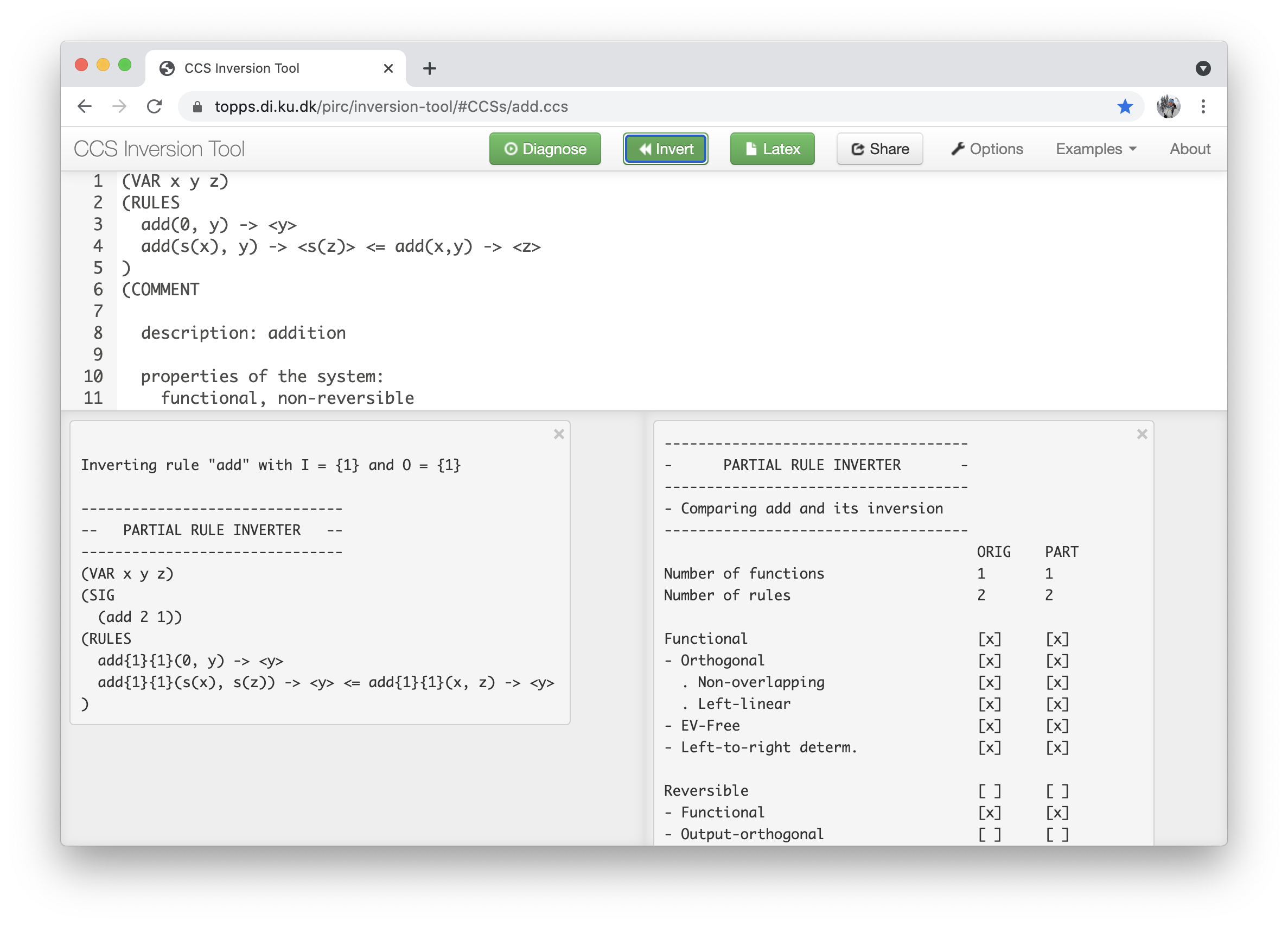}
\end{center}
\vspace{-3ex}
\caption{The interface of the web-based inversion tool after partially inverting \texttt{add} to \texttt{add\{1\}\{1\}}.}
\label{fig:ack-screenshoot}
\end{figure}

Whereas the source code provides a
command line interface, which facilitates composition with other program transformations, the online web-based version provides a friendly clickable interaction; see Figure~\ref{fig:ack-screenshoot} for a screenshot. In the following, we describe the most important content and features of the online tool.
The tool web-site contains the following:
\begin{enumerate}
\item a navigation bar (in the top) with green action buttons and white settings buttons,
\item a white input window with a text field for the original CCS,
\item a gray output window (in the lower left corner) for the inverted systems,
and
\item another gray output window (in the lower right corner) with program diagnostics.
\end{enumerate}
The original CCS can either be entered into the input window or chosen from the predefined CCS examples available via the \emph{Examples} button, e.g., choosing \texttt{add}.
Using the \emph{Options} button, one defines the inversion task, e.g., the partial inversion of \texttt{add} with
$I=\{1\}$ and $O=\{1\}$ and selects one of the rule inverters, e.g., the partial rule inverter.
To apply the inverter, we use the
\emph{Invert} button whereafter the tool creates (or updates) the gray output windows with the inversion, e.g., the partial inversion \texttt{add\{1\}\{1\}}, and the paradigm characteristics of both the original and the inverted program, e.g., column \texttt{ORIG} and column \texttt{PART}.
For instance, we can see that both \texttt{add} and \texttt{add\{1\}\{1\}} are functional and that none of them is reversible.
The
\emph{Diagnose} button provides a more detailed property analysis of the program in the white input text field.
Another feature is the
\emph{Latex} button that translates the CCS in the main window into
{\LaTeX}
code that can be used when typesetting documents.

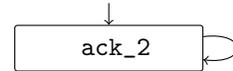
\begin{figure}[t]
\captionsetup[subfigure]{aboveskip=-5pt,belowskip=5pt}
\begin{center}
\begin{subfigure}[t]{\textwidth}
\begin{minipage}[c][2.5cm]{0.8\textwidth}
\small
\begin{verbatim}
ack(0,y)       -> <s(y)>
ack(s(x),0)    -> <z> <= ack(x,s(0)) -> <z>
ack(s(x),s(y)) -> <z> <= ack(s(x),y) -> <v>, ack(x,v) -> <z>\end{verbatim}
\end{minipage}
\begin{minipage}[c][2.5cm]{0.15\textwidth}
\begin{tikzpicture}
\def\dl{240} 
\def\dr{300} 
\def\ul{120} 
\def\ur{60}  

    \node[shape=rectangle,rounded corners=1pt,draw=black,minimum width=2.5cm] (ack) at (0,1) {\small{\texttt{
    \phantom{\{}
    ack
    \phantom{\}}
    }}};

    \path[->] (0,1.6) edge (ack);
    
    \path [->] (ack)     edge [out=7, in=-7, loop,looseness=5]        (ack);

\end{tikzpicture}
\end{minipage}
\caption{Program \texttt{ack} implementing the Ackermann function $\mathit{Ack}(x,y)$.}
\label{fig:ack}
\end{subfigure}
\begin{subfigure}[t]{\textwidth}
\begin{minipage}[c][4cm]{0.8\textwidth}
\small
\begin{verbatim}
ack{1}{1}(0,s(y)) -> <y>
ack{1}{1}(s(x),z) -> <0>    <= ack{1,2}{1}(x,s(0),z) -> <>
ack{1}{1}(s(x),z) -> <s(y)> <= ack{1}{1}(x,z)    -> <v>,
                               ack{1}{1}(s(x),v) -> <y>
ack{1,2}{1}(0,y,s(y))    -> <>
ack{1,2}{1}(s(x),0,z)    -> <> <= ack{1,2}{1}(x,s(0),z) -> <>
ack{1,2}{1}(s(x),s(y),z) -> <> <= ack{1}{1}(x,z) -> <v>, 
                                  ack{1,2}{1}(s(x),y,v) -> <>\end{verbatim}
\end{minipage} \begin{minipage}[c][4cm]{0.14\textwidth}
\begin{tikzpicture}
\def\dl{240} 
\def\dr{300} 
\def\ul{120} 
\def\ur{60}  

    \node[shape=rectangle,rounded corners=1pt,draw=black,minimum width=2.5cm] (ack121) at (0,0) {\small{\texttt{ack\{1,2\}\{1\}}}};
    \node[shape=rectangle,rounded corners=1pt,draw=black,minimum width=2.5cm] (ack11) at (0,1) {\small{\texttt{ack\{1\}\{1\}}}};

    \path[->] (0,1.6) edge (ack11);
    
    \path [->](ack121.\ur) edge[bend right=30]     (ack11.\dr);
    \path [->] (ack121)     edge [out=7, in=-7, loop,looseness=5]        (ack121);

    \path [->](ack11.\dl)   edge[bend right=30]     (ack121.\ul);
    \path [->](ack11)       edge [out=7, in=-7,loop,looseness=5]       (ack11);

\end{tikzpicture}
\end{minipage}                                
\caption{Partial inverse of \texttt{ack} with $I=\{1\}$ and $O=\{1\}$.} 
\label{fig:ack-part}
\end{subfigure}
\begin{subfigure}[t]{\textwidth}
\begin{minipage}[c][2.5cm]{0.8\textwidth}
\small
\begin{verbatim}
ack_2(0,s(y)) -> <y>
ack_2(s(x),z) -> <0>    <= ack_2(x,z) -> <s(0)>
ack_2(s(x),z) -> <s(y)> <= ack_2(x,z) -> <v>,
                           ack_2(s(x),v) -> <y>\end{verbatim}
\end{minipage}
\begin{minipage}[c][2.5cm]{0.15\textwidth}
\begin{tikzpicture}
\def\dl{240} 
\def\dr{300} 
\def\ul{120} 
\def\ur{60}  

    \node[shape=rectangle,rounded corners=1pt,draw=black,minimum width=2.5cm] (ack2) at (0,1) {\small{\texttt{
    \phantom{\{}
    ack\_2
    \phantom{\{}
    }}};

    \path[->] (0,1.6) edge (ack2);
    
    \path [->] (ack2)     edge [out=7, in=-7, loop,looseness=5]        (ack2);

\end{tikzpicture}
\end{minipage}
\caption{Romanenko's partial inverse $\mathit{Ack}_2^{-1}$~\cite[p.17]{Romanenko1991} rewritten as a CCS.}
\label{fig:ack-rom}
\end{subfigure}
\end{center}

\caption{Partial inversions of the Ackermann function and the dependency graphs.
}
\label{fig:ack-org-inv}
\end{figure}

\section{A Case Study of Ackermann Inversion}
\label{sec:ackermann}

We illustrate the use of our tool by repeating three experiments~\cite{Romanenko1991}, namely, two partial inversions and a full inversion of the Ackermann function \texttt{ack} (Figure~\ref{fig:ack}). \texttt{ack} takes two unary numbers as inputs and returns one unary number as output.
\paragraph{First Experiment}
An io-set together with the \texttt{ack} program in Figure~\ref{fig:ack} are the tool inputs.
The io-set for this experiment is $I=\{1\}$ and $O=\{1\}$, specifying that the first input term and the output term of \texttt{ack} are the inputs for the partially inverted program \texttt{ack\{1\}\{1\}}.
Then, our tool propagates the io-set through the entire program and transforms the rules locally using the selected well-behaved rule inverter.

The result of the pure partial inversion~\cite[Fig.6]{KirGlu:20} is shown in Figure~\ref{fig:ack-part}.
The resulting program consists of two defined function symbols, namely, the desired partial inverse \texttt{ack\{1\}\{1\}}, which depends on another partial inverse \texttt{ack\{1,2\}\{1\}}.
The io-set of \texttt{ack\{1,2\}\{1\}} specifies that it takes both inputs and the output of \texttt{ack} as the input. As a consequence, all of its three rules return
a nullary output tuple~\texttt{<>}.
In the case of Romanenko's \texttt{ack\_2} in Figure \ref{fig:ack-rom}, the second rule's right-hand side of the condition is a constant \texttt{s(0)}. Since this output is a known constant, we can provide it as input to the left-hand side using our tool.
This illustrates that our tool fully propagates the io-sets such that all known terms become the new input.
This means that the algorithm is a \emph{polyvariant inverter} in that
it may produce several inversions of the
same function symbol, namely, one for each input-output index set.

The relation specified by the partial inverse
is functional~\cite[p.18]{Romanenko1991},
but the program in Figure~\ref{fig:ack-part} is nondeterministic
due to a single pair of overlapping rules, i.e.,\ the 2nd and 3rd rules of \texttt{ack\{1\}\{1\}}.
The same issue occurs for the partially inverted program {$\mathit{Ack}_2^{-1}$}
(\texttt{ack_2}, Figure~\ref{fig:ack-rom})~\cite[p.17]{Romanenko1991}.
Comparison of the two programs shows that in \texttt{ack\{1\}\{1\}}'s second rule, our tool has moved the condition's constant \texttt{s(0)} to the input side
and thereby created a dependency on the more specific partial inversion \texttt{ack\{1,2\}\{1\}} instead of \texttt{ack\{1\}\{1\}}.

The effect is to reduce the search space when rewriting
using the inverted systems.
In this experiment, we found a remarkable reduction
of function calls and rewrite steps.
The results and the speed-ups for \texttt{ack_2} and \texttt{ack\{1\}\{1\}} are reported in Table~\ref{tab:ack2ack11evaluation}.
For the tested inputs ranging from \mbox{(1, 2)} to \mbox{(3, 509)}
the speed-up is up to 5.89 for rewrite steps and up to 3.92 for function calls when comparing \texttt{ack\{1\}\{1\}} with Romanenko's \texttt{ack_2}.
We observe that there is a rewrite step speed-up for all inputs, while the two programs have equally many function calls when the first input 
is~1.
One reason is that \texttt{ack\{1\}\{1\}} tends to use fewer rewriting steps because its call to \texttt{ack\{1,2\}\{1\}} can fail using pattern matching, whereas \texttt{ack_2} requires a rewriting and pattern match of the result to establish the same~failure.

The number of required rewrite steps is used in the complexity of conditional term rewriting systems~\cite{Kop2017}, and the number of function/predicate calls is used in the complexity analysis of functional and logic programs~\cite{Metayer1988,Navas2007}. Here, function calls correspond to
the number of function-rooted terms that must be rewritten to reach normal form.

\begin{table}[t]
\footnotesize \begin{center}
\begin{tabular}{@{}llrrrrrrr@{}}
\toprule
 Input & & (1, 2) & (1, 3) & (1, 4) & (1, 5) & (1, 6) & (1, 7) & (1, 8) \\ \midrule
 \texttt{ack\_2} & Rewrite steps        & 5      & 8      & 11     & 14     & 17     & 20     & 23     \\
& Function calls        & 9      & 12     & 15     & 18     & 21     & 24     & 27  \\ \midrule
 \texttt{ack\{1\}\{1\}} & Rewrite steps & 4      & 6      & 8      & 10     & 12     & 14     & 16     \\
 & Function calls & 9      & 12     & 15     & 18     & 21     & 24     & 27  \\ \midrule

\textit{Speed-up} & Rewrite steps  & 1.25   & 1.33    & 1.38    & 1.40    & 1.42     & 1.43     & 1.53  \\
& Function calls & 1.00   & 1.00    & 1.00    & 1.00    & 1.00     & 1.00     & 1.00    \\ \bottomrule                   

\\[5pt]

\toprule
   Input                 & & (2, 3) & (2, 5) & (2, 7) & (2, 9) & (2, 11) & (2, 13) & (2, 15) \\ \midrule
 
 \texttt{ack\_2} & Rewrite steps        & 21     & 50     & 91     & 144    & 209     & 286     & 375  \\
&Function calls        & 38     & 75     & 124    & 185    & 258     & 343     & 440  \\ \midrule

\texttt{ack\{1\}\{1\}} & Rewrite steps & 13     & 25     & 41     & 61     & 85      & 113     & 145  \\
             &Function calls & 28     & 51     & 80     & 115    & 156     & 203     & 256 \\ \midrule
\textit{Speed-up}& Rewrite steps  & 1.62   & 2.00    & 2.22    & 2.36    & 2.46     & 2.53     & 2.59  \\
& Function calls & 1.36   & 1.47    & 1.55    & 1.61    & 1.65     & 1.69     & 1.72        
             \\ \bottomrule
\\[5pt]

\toprule
     Input        & & (3, 5) & (3, 13) & (3, 29) & (3, 61) & (3, 125) & (3, 253) & (3, 509) \\ \midrule
\texttt{ack\_2}& Rewrite steps        & 109    & 682     & 3351    & 14820   & 62321    & 255614   & 1035403  \\
 & Function calls        & 178    & 865     & 3776    & 15743   & 64254    & 259581   & 1043452  \\ \midrule
\texttt{ack\{1\}\{1\}}& Rewrite steps & 45     & 186     & 727     & 2836    & 11153    & 44174    & 175755  \\
& Function calls & 95     & 347     & 1239    & 4563    & 17359    & 67531    & 266183   \\ \midrule
\textit{Speed-up} & Rewrite steps  & 2.42   & 3.67    & 4.61    & 5.23    & 5.59     & 5.79     & 5.89  \\
& Function calls & 1.87   & 2.49    & 3.05    & 3.45    & 3.70     & 3.84     & 3.92         \\ \bottomrule             
\end{tabular}
\end{center}

\caption{
The rewrite steps and function calls for \texttt{ack\{1\}\{1\}} and \texttt{ack\_2} on a range of inputs.}
\label{tab:ack2ack11evaluation}
\end{table}

To confirm that these speed-ups
manifest themselves in a functional-logic language,
we implemented \texttt{ack\{1\}\{1\}} and \texttt{ack_2}
in Curry and measured their runtimes in CPU seconds on input (3,253)
using two Curry systems\footnote{The programs were executed using the Docker images \texttt{caups/pakcs3:3.3.0} and \texttt{caups/kics2:2.3.0} on an Apple MacBook Pro (2.6 GHz 6-Core Intel Core i7 processor, 16 GB memory, Intel Graphics). The execution times are slower than if Curry were installed directly on the machine, but the relative program execution times are expected to hold in either case.}:
The Haskell-based Kics2 terminated on the programs after 1295.7~s and 4674.6~s, respectively, and the Prolog-based Pakcs3 terminated after 8.5~s and 40.7~s, respectively. Thus, the speed-ups in Curry, which are
3.61 and 4.79, are comparable to the speed-ups for function calls and rewrite~steps.

On the other hand, the polyvariant io-set propagation also has a cost with respect to the size of \texttt{ack\{1\}\{1\}}: in the worst case, all possible inversions of a function symbol are created--io-sets are never generalized--thereby increasing the size of the generated program. Despite the full propagation of the io-sets, the tool always terminates due to their finite number for any program; this characteristic relates to mode analysis~\cite{Ohlebusch2002}.
Romanenko's method, which is potentially more powerful due to the global approach because it builds
a configuration graph
and uses generalization to make the unfolding of calls terminate, produces a \emph{monovariant} partial inverse \texttt{ack_2}
so that
not all known local information is used
(Figure~\ref{fig:ack-rom}); this may be due to the generalization in the configuration graph~\cite{Romanenko1988}.

\begin{figure}[t]
\captionsetup[subfigure]{aboveskip=-5pt,belowskip=5pt}
\begin{center}
\begin{minipage}[c]{\textwidth}
\begin{subfigure}[t]{\textwidth}
\begin{center}
\begin{minipage}[b][4cm]{0.65\textwidth}
\small
\begin{verbatim}
ack{2}{1}(y, s(y)) -> <0>
ack{2}{1}(0, z) -> <s(x)>    <= ack{2}{1}(s(0), z) -> <x>
ack{2}{1}(s(y), z) -> <s(x)> <= ack{}{1}(z) -> <x, v>, 
                                ack{1,2}{1}(s(x), y, v) -> < >
ack{}{1}(s(y)) -> <0, y>
ack{}{1}(z) -> <s(x), 0>    <= ack{2}{1}(s(0), z) -> <x>
ack{}{1}(z) -> <s(x), s(y)> <= ack{}{1}(z) -> <x, v>, 
                               ack{1}{1}(s(x), v) -> <y>
\end{verbatim}
\end{minipage}
\end{center}
\caption{
Partial inverse of \texttt{ack} with $I=\{2\}$ and $O=\{1\}$ includes, in addition, the rules of Figure~\ref{fig:ack-part}.
}
\label{fig:ack-part-two}
\end{subfigure}
\begin{subfigure}[t]{\textwidth}
\begin{center}
\begin{minipage}[t]{0.6\textwidth}
\small
\begin{verbatim}
ack_1(y, s(y)) -> <0>
ack_1(0, z) -> <s(x)>    <= ack_1(s(0), z) -> <x>
ack_1(s(y), z) -> <s(x)> <= ack_0(z) -> <s(x), v>, 
                            ack_1(s(y), v) -> <s(x)>

ack_0(s(y)) -> <0, y>
ack_0(z) -> <s(x), 0>    <= ack_0(z) -> <x, s(0)>
ack_0(z) -> <s(x), s(y)> <= ack_0(z) -> <x, v>, 
                            ack_0(v) -> <s(x), y>  \end{verbatim}
\end{minipage}
\end{center}
\caption{Romanenko's partial inverse $\mathit{Ack}_1^{-1}$~\cite[p.17]{Romanenko1991} rewritten as a CCS.}
\label{fig:ack-rom-two}
\end{subfigure}
\begin{center}
\begin{subfigure}[t]{0.8\textwidth}
\begin{minipage}[t][3cm]{\textwidth}
\begin{center}
\begin{minipage}[t][3cm]{0.6\textwidth}
\begin{center}
\begin{tikzpicture}
\def\dl{240} 
\def\dr{300} 
\def\ul{120} 
\def\ur{60}  

    \node[shape=rectangle,rounded corners=1pt,draw=black,minimum width=2.5cm] (ack21) at (0,1) {\small\texttt{ack\{2\}\{1\}}};
    \node[shape=rectangle,rounded corners=1pt,draw=black,minimum width=2.5cm] (ack1) at (0,0) {\small{\texttt{ack\{\}\{1\}}}};
    \node[shape=rectangle,rounded corners=1pt,draw=black,minimum width=2.5cm] (ack121) at (3,1) {\small{\texttt{ack\{1,2\}\{1\}}}};
    \node[shape=rectangle,rounded corners=1pt,draw=black,minimum width=2.5cm] (ack11) at (3,0) {\small{\texttt{ack\{1\}\{1\}}}};
    
    \path[->] (0,1.6) edge (ack21);

    \path [->](ack21.\dl)  edge[bend right=30]     (ack1.\ul);
    \path [->](ack21)       edge                    (ack121);
    \path [->](ack21)       edge [out=173, in=187, loop,looseness=5]        (ack21);
    
    \path [->](ack1.\ur)    edge[bend right=30]     (ack21.\dr);
    \path [->](ack1)        edge                    (ack11);
    \path [->] (ack1) edge [ out=173, in=187, loop,looseness=5]       (ack1);

    \path [->](ack121.\dl) edge[bend right=30]     (ack11.\ul);
    \path [->] (ack121)     edge [out=7, in=-7, loop,looseness=5]        (ack121);

    \path [->](ack11.\ur)   edge[bend right=30]     (ack121.\dr);
    \path [->](ack11)       edge [out=7, in=-7,loop,looseness=5]       (ack11);

\end{tikzpicture}
\end{center}
\end{minipage}
\begin{minipage}[t][3cm]{0.35\textwidth}
\begin{center}
\begin{tikzpicture}
\def\dl{240} 
\def\dr{300} 
\def\ul{120} 
\def\ur{60}  

    \node[shape=rectangle,rounded corners=1pt,draw=black,minimum width=2.5cm] (ack1) at (0,1) {\small{\texttt{
    \phantom{\{}
    ack\_1 
    \phantom{\{}
    }}};
    \node[shape=rectangle,rounded corners=1pt,draw=black,minimum width=2.5cm] (ack0) at (0,0) {\small{\texttt{
    \phantom{\{}
    ack\_0 \phantom{\{}}}};

    \path[->] (0,1.6) edge (ack1);
    
    \path [->](ack1.\dl) edge[bend right=30]     (ack0.\ul);
    \path [->] (ack1)     edge [out=7, in=-7, loop,looseness=5]        (ack1);

    \path [->](ack0)       edge [out=7, in=-7,loop,looseness=5]       (ack0);

\end{tikzpicture}
    
\end{center}
\end{minipage}
\end{center}
\end{minipage}
\vspace{-3ex}
\caption{Dependency graphs of \texttt{ack\{2\}\{1\}} and \texttt{ack\_1}.}
\label{fig:ack-inv-two-dep-graphs}
\end{subfigure}

\end{center}

\end{minipage}
\end{center}
\caption{A partial inversion of the Ackermann function and the dependency graphs.
}
\label{fig:ack-inv-two}
\end{figure}
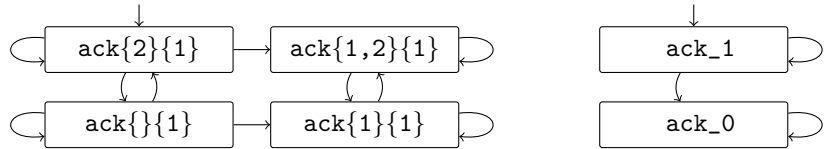

\paragraph{Second experiment}
The next experiment is the
partial inversion \texttt{ack\{2\}\{1\}} and our tool correctly produces the inverse that defines four function symbols including \texttt{ack\{1\}\{1\}} and \texttt{ack\{1,2\}\{1\}} and also a full inverse \texttt{ack\{\}\{1\}}. This full inverse depends on the partial inverses \texttt{ack\{1\}\{1\}}
and \texttt{ack\{2\}\{1\}}
due to
the io-set propagation in our tool.
By contrast, Romanenko's partial inversion \mbox{\texttt{ack\_1}}
depends on itself and on \texttt{ack_1}'s full inverse \texttt{ack_0}.
His full inverse depends on itself~\cite[p.17]{Romanenko1991} instead of partial inverses that would have been possible if all known information was exploited.
Both systems \texttt{ack\{2\}\{1\}} and \texttt{ack_1} are shown in Figure~\ref{fig:ack-part-two} and~\ref{fig:ack-rom-two}, where \texttt{ack\{2\}\{1\}} depends on \texttt{ack\{1,2\}\{1\}} and \texttt{ack\{1\}\{1\}} in Figure~\ref{fig:ack-part}.
The systems are illustrated by their dependency graphs in Figure~\ref{fig:ack-inv-two-dep-graphs}.

Romanenko's \texttt{ack_1} and \texttt{ack\{2\}\{1\}} are nonterminating.
The third rule of \texttt{ack\_0} has \texttt{ack\_0(z)} as its left-hand side and  also requires a rewriting of the same term \texttt{ack\_0(z)} in its first condition, thus yielding an infinitely deep search tree. The third rule of \texttt{ack\{\}\{1\}} has a similar structure. Since both programs are nonterminating, no counts are provided.
Nevertheless, when producing \texttt{ack\{2\}\{1\}}, our tool discovers an improvement of the inverse system, e.g., the second condition of the third rule depends on the terminating \texttt{ack\{1,2\}\{1\}}
whereas  the same condition of the same rule of \texttt{ack\_1} depends on the nonterminating \texttt{ack\_1}.

The cost of creating polyvariant inversions is evident in \texttt{ack\{2\}\{1\}}, where the tool has created 4 different inversions of the 3 original rules, producing a system of 12 rules. In comparison, \texttt{ack_1} consists of two inversions of the same three original rules producing a smaller system of 6 rules; see the dependency graphs in Figure~\ref{fig:ack-inv-two-dep-graphs}.

\begin{figure}[t]
\begin{center}
\begin{minipage}[c]{0.65\textwidth}
\small
\begin{verbatim}
ack{}{1}(s(y)) -> <0, y>
ack{}{1}(z) -> <s(x), 0> <= ack{}{1}(z) -> <x, s(0)>
ack{}{1}(z) -> <s(x), s(y)> <= ack{}{1}(z) -> <x, v>,
                               ack{}{1}(v) -> <s(x), y>\end{verbatim}
\end{minipage}
\end{center}
\vspace{-3ex}
\caption{The full inverse of the Ackermann function.}
\label{fig:ack-inv-full}
\end{figure}

\paragraph{Third experiment}
In the third experiment, Romanenko used his full inverter~\cite[Sect.3.1]{Romanenko1991} to invert \texttt{ack}, and our pure full inverter~\cite[Fig.6]{KirGlu:20} produces exactly the same program, namely, \texttt{ack\{\}\{1\}}, in Figure~\ref{fig:ack-inv-full}. 
Please note that this full inversion shares the same defined function symbol as the rules in Figure~\ref{fig:ack-inv-two}, but the rules are different.
This is because they define the same input-output relation, namely, 
the full inversion of the original \texttt{ack}.
This system is nonterminating; thus, no count is provided.
By exploiting the mathematical property of Ackermann that its output is larger than its input, it may be possible to create a terminating full inversion. It is beyond the tool to use extra mathematical properties to improve the inversions.

The fourth partial inversion that is possible is \texttt{ack\{1,2\}\{1\}}, which is already included in Figure~\ref{fig:ack-part}. This means that with our tool, we produced all four possible partial inversions (including the special case of full inversion) of the Ackermann function in the course of the three experiments.
Using our tool, we also reproduced all of the examples in~\cite{KirGlu:20,KirGlu:19}.

\section{Conclusion and Future Work}
\label{sec:conclusion}
The goal of this work was to provide a design space for the experimental evaluation and comparison of different well-behaved rule inverters, including those using heuristic approaches~\cite{KirGlu:19}. It will be interesting to investigate Romanenko's inversion method~\cite{Romanenko1988} as well as related global approaches~\cite{Gluck2005,GlueckTurchin:90,Kawabe2005} and program analyses such as mode and binding-time analyses.
Using CCSs enabled us to focus on the essence of inversion without considering language-specific details, as demonstrated by the examples above.
The examples demonstrate that polyvariant inversion can considerably reduce the search space of the inverted system.
The post-optimizations of the inverted programs represent another future direction of investigation.
We have observed two potential improvements: the first is the reduction of nondeterminism by determinization~\cite{Kawabe2005,Nagashima2012},
and the other is exploiting constants by partial evaluation, for example, the constant \texttt{s(0)} of the 2nd rule of Figure~\ref{fig:ack-part}.
We expect that this will further improve the efficiency of inverse systems.
In future work, one can consider the translation of the resulting programs to logic or functional-logic programming languages, such as
Prolog or Curry, and
explore the relation to partial deduction in logic programming.


\paragraph{Acknowledgements} 
Thanks to Alberto Pettorossi and to the anonymous reviewers for their constructive feedback on an earlier version
of this paper.


\bibliographystyle{eptcs}
\bibliography{generic}

\end{document}